\documentclass[doublecolumn]{mn2e}
\usepackage{graphicx}
\begin{document}
\title[$R$-mode oscillations of rapidly rotating relativistic stars]{
$R$-mode oscillations of rapidly rotating barotropic stars in general
relativity: Analysis by the relativistic Cowling approximation}
\author[S. Yoshida, S'i.Yoshida and Y. Eriguchi]{
Shijun Yoshida$^{1}$\thanks{E-mail: shijun@waseda.jp},
Shin'ichirou Yoshida$^{2}$\thanks{E-mail: sy@uwm.edu}
and
Yoshiharu Eriguchi$^3$\thanks{E-mail: eriguchi@esa.c.u-tokyo.ac.jp} \\
$^1$Science and Engineering, Waseda University, Okubo, Shinjuku,
Tokyo 169-8555, Japan \\
$^2$Department of Physics, University of Wisconsin-Milwaukee,
1900 E. Kenwood Blvd., Milwaukee, WI  53211, USA \\
$^3$Department of Earth Science and Astronomy, Graduate School of Arts
and Sciences, University of Tokyo, \\
Komaba, Meguro, Tokyo 153-8902, Japan
}
\date{Typeset \today ; Received / Accepted}
\maketitle
\begin{abstract}

We develop a numerical scheme for obtaining the $r$-mode oscillations
of rapidly rotating relativistic stars. In the present scheme, we neglect
all metric perturbations and only take account of the dynamics of the
fluid in the background spacetime of the unperturbed star (the
relativistic Cowling approximation). We also assume the star is barotropic,
i.e. neutrally stable against convection under the assumption of adiabatic
oscillations. Our  numerical scheme is based on the Yoshida-Eriguchi
formulation for the analysis of the general relativistic f-mode oscillations
in the Cowling approximation and a general relativistic generalization
of the Karino-Yoshida-Yoshida-Eriguchi's numerical scheme for obtaining
oscillations of rapidly rotating Newtonian stars. By this new numerical
scheme, frequencies of the $r$-mode oscillations are obtained as functions of
the ratio of the rotational energy to the absolute value of the
gravitational energy $T/|W|$ along sequences of polytropic equilibrium
stars whose ratio of the pressure to the total energy density at the center
of the star and the polytropic index are kept constant. It is found that the
dimensionless oscillation frequency $\sigma/\Omega$ is a linearly decreasing
function of $T/|W|$, where $\sigma$ and $\Omega$ are the oscillation
frequency and the angular velocity of the star measured in an inertial frame
at spatial infinity. We also find that oscillation frequencies of the
$r$-modes are highly dependent on the relativistic factor $M/R$ of the star
as already found in previous studies in which the slow rotation approximation
has been used. Here $M$ and $R$ denote the mass and radius of the star,
respectively.

\end{abstract}

\section{Introduction}

The inertial modes as well as the $f$-modes are basic oscillation modes of
rotating stars in the sense that they exist even in
incompressible stars (Bryan 1889; Lindblom \& Ipser 1999). The inertial mode
oscillations are mainly restored by the Coriolis force due to stellar
rotation and their oscillation frequencies are therefore comparable to
the stellar
rotation frequencies (Papaloizou \& Pringle 1978; Provost, Berthomieu
\& Rocca 1981; Saio 1982; Unno et al. 1989). The $r$-mode oscillations
belong to a sub-class of the inertial mode oscillations and are characterized
by small radial displacement vectors of the eigenfunctions. They
are unstable against gravitational radiation reactions for all
rotating stars if the stars are inviscid ($r$-mode instability:
Andersson 1998; Friedman \& Morsink 1998). The $r$-mode instability is
attributed to the so-called Chandrasekhar-Friedman-Schutz (CFS) mechanism,
which makes various modes of oscillations of rotating stars unstable
(Chandrasekhar 1970; Friedman \& Schutz 1978). The $r$-mode
oscillations excited in neutron stars due to the CFS mechanism may
slow the rotation of newly born
neutron stars as well as cold old neutron stars in low mass x-ray binary
(LMXB) systems. Furthermore, there have been suggestions that
the gravitational radiation emitted due to the
excitation of the $r$-mode oscillations will be observed by the Laser
Interferometric Gravitational Wave Observatory (LIGO) and other detectors
(Lindblom, Owen \& Morsink 1998; Andersson, Kokkotas \& Schutz 1999; Owen
et al. 1998; Bildsten 1998; Andersson, Kokkotas \& Stergioulas 1999;
See also the review by Andersson and Kokkotas 2001), though the non-linear
saturation of the mode may inhibit the mechanism to work in newly
born stars (Arras et al. 2003).

So far, most studies of the $r$-mode oscillations of neutron stars have been
carried out within the slow rotation approximation and/or the weak
gravitational field approximation (Newtonian dynamics). However, analyses
under these approximations are not sufficient to investigate the $r$-mode
oscillations of neutron stars because of the following reasons. First, the
effect of general relativity on the $r$-mode oscillations of neutron stars
would be significant because the relativistic factor of neutron stars, $M/R$,
could be as large as $M/R\sim 0.2$, where $M$ and $R$ denote the mass and
radius of the star, respectively.
Since the effect of dragging of inertial frame on the angular velocity
of the fluid measured by the locally non-rotating observer is in this
order, the inertial mode should be also modified to this order
(see Eq.(\ref{localw}) and the discussion there).
Second, from time to time, the $r$-mode
instability of nascent neutron stars has been investigated because this is
one of the most important issues of the $r$-mode instability related to
neutron stars. The spin rates of neutron stars at their birth are believed
to be near the mass-shedding limit where the mass of the star begins to
shed from the stellar surface of the equator. If nascent neutron stars are
considered, therefore, the effect of deformation of the star due to rapid
stellar rotation or of the centrifugal force on the $r$-mode oscillations
should not be neglected. In short, both effects of general relativity and
rapid rotation have to be taken into account in studies of the $r$-mode
oscillations of neutron stars.

The $r$-mode oscillations of rapidly rotating compressible stars have
been obtained for the first time by Yoshida et al. (2000) for uniformly
rotating stars within the framework of Newtonian gravity. They have shown
that the basic properties of the $r$-mode oscillations of slowly rotating
stars do not change greatly and that extrapolating the results of the slow
rotation approximation gives us good estimates for properties of the
$r$-mode oscillations and the $r$-mode instability even for the rapidly
rotating models (see, also, Karino et al. 2000). The Newtonian scheme has
been applied to differentially and rapidly rotating stars by Karino,
Yoshida \& Eriguchi (2001).

As for general relativistic investigations of the $r$-mode oscillations,
Kojima (1998) considered for the first time the axial parity perturbations
of slowly rotating stars by assuming the oscillation frequencies to be
proportional to the stellar angular velocities (low-frequency
approximation) and derived a single second order ordinary differential
equation (Kojima's equation) for the $r$-mode-like oscillations. He found
that the relativistic $r$-mode oscillations are absolutely different from
Newtonian ones in the sense that Kojima's equation allows a continuous
spectrum for the eigenfrequency (see, also, Beyer \& Kokkotas 1999).
Lockitch, Andersson \& Friedman (2001) noticed that Kojima's equation is
valid only if stars are non-barotropic and that, for barotropic stars,
the polar parity perturbations have to be taken into account as well as
the axial parity perturbations in order to obtain the $r$-mode-like
oscillations. For barotropic stars, they then obtained a general relativistic
counterpart of the $r$-mode oscillation in Newtonian stars. Their
eigenfunctions are composed both of the polar and axial parity perturbations
(see, also, Lockitch 1999; Lockitch, Friedman \& Andersson 2003).
Their analysis has been recently extended to superfluid stars by
Yoshida \& Lee (2003).

For the relativistic $r$-mode oscillations of non-barotropic stars, the
situation is different from that of barotropic stars. For non-barotropic
stars, as mentioned before, if we use the low-frequency approximation and the
slow rotation approximation, basic equations becomes Kojima's equation, which
contains a singular point inside the star for some frequency range and
yields a continuous spectrum and singular eigenfunctions (Kojima 1998).
The main issue, which has not been fully solved yet, is whether there
exist the $r$-mode-like solutions characterized by discrete eigenfrequencies
and regular eigenfunctions in non-barotropic relativistic stars or not.
In spite of various attempts to improve our understanding of the relativistic
$r$-mode oscillations, we have not yet understood the $r$-mode oscillations
of non-barotropic relativistic stars clearly
(Kojima \& Hosonuma 1999; Kojima \& Hosonuma 2000;
Lockitch, Andersson \& Friedman 2001; Yoshida 2001; Ruoff \& Kokkotas
2001; Yoshida \& Futamase 2001; Ruoff \& Kokkotas 2002; Yoshida \& Lee
2002; Ruoff, Stavridis \& Kokkotas 2003). The present study focuses upon
the $r$-mode oscillations of barotropic stars. We will study non-barotropic
stars in a forthcoming paper.

The main purpose of this paper is to improve our understanding of properties
of the $r$-mode oscillations of rapidly rotating relativistic stars.
Particularly, we want to explore both the effects of general relativity and
rapid rotation on the $r$-mode oscillations. Up to now, there has been no
prescription for managing pulsations of rapidly rotating relativistic stars
because of difficulties due to gauge conditions, boundary conditions at
spatial infinity and other factors, although there is a method to obtain zero
frequency $f$-modes devised by Stergioulas \& Friedman (1998). However,
this method is not available for obtaining general oscillation modes of
non-zero frequencies. In order to avoid these difficulties,
we employ a relativistic version of the Cowling approximation in which all
the metric perturbations are neglected (McDermott, Van Horn \& Scholl 1983;
Finn 1988; Ipser \& Lindblom 1992; Yoshida \& Eriguchi 1997). As discussed
by Lindblom \& Splinter (1990), Yoshida \& Kojima (1997) and
Yoshida \& Eriguchi (1997), the relativistic Cowling approximation
is a good approximation for determining frequencies of oscillation modes
for which the fluid motions dominate vibrations of the spacetime.
Therefore, the Cowling approximation would be suitable for analyzing
the $r$-mode oscillations for which the fluid motions are dominating.

Making use of the relativistic Cowling approximation, we calculate
the $r$-mode oscillations of rapidly rotating stars. In this paper, we
assume oscillations to be adiabatic and consider only barotropic stars
as mentioned before. In order to obtain the $r$-mode oscillations of rapidly
rotating relativistic stars, we make use of the relativistic Cowling
formulation for the rapidly rotating polytropes by Yoshida \& Eriguchi (1997)
and generalize  to rapidly rotating general relativistic stars
the Karino-Yoshida-Yoshida-Eriguchi numerical method
(Karino et al. 2000) for obtaining the $r$-mode oscillations of rapidly
rotating Newtonian stars.
The generalization is straightforward. Since we employ the relativistic
Cowling approximation, basic equations are those of fluid motions and very
similar to those of the Newtonian stars (we do not need to solve the
linearized
Einstein equation). In \S 2 we present the basic equations and the numerical
method employed in this paper for studying pulsations of rapidly rotating
relativistic stars. Note that notations are not the same as those of
Yoshida and Eriguchi (1997). Numerical results are given in \S 3, and \S 4 is
devoted to discussion and summary. In this paper, we use units in which
$c=G=1$, where $c$ and $G$ denote the velocity of light and the gravitational
constant, respectively.

\section{Formulation}

\subsection{Equilibrium State}

We consider oscillations of axisymmetric equilibrium configurations of
uniformly rotating polytropic stars without meridional circulations. The
spacetime for such configurations can be described by the line element:
\begin{eqnarray}
ds^2=-e^{2\nu}\,dt^2&+&e^{2\alpha}\,(dr^2+r^2d\theta^2)\nonumber\\
&+&e^{2\beta}\,r^2\sin^2\theta\,(d\varphi-\omega\,dt)^2\,,
\end{eqnarray}
where $\nu$, $\alpha$, $\beta$, and $\omega$ are functions of $r$ and
$\theta$ only. Here spherical coordinates $(r, \theta, \varphi)$
are used. The four velocity of the fluid, $u^\mu$, is given by
\begin{eqnarray}
u^\mu= \gamma (t^\mu+\Omega\,\varphi^\mu)\,,
\end{eqnarray}
\begin{eqnarray}
 \gamma = {e^{-\nu} \over \sqrt{1 - v^2}} \ ,
\end{eqnarray}
\begin{eqnarray}
 v = r \sin \theta e^{\beta - \nu} (\Omega - \omega) \ ,
\end{eqnarray}
where $t^\mu$ and $\varphi^\mu$ are, respectively, the time-like and
rotational Killing vectors, and $\Omega$ is the constant angular velocity
of the star. The polytropic equation of state is defined by
\begin{eqnarray}
 p= K \,\rho^{1+1/N}\,,\quad
\varepsilon=\rho+N\,p\,,
\end{eqnarray}
where $p$, $\rho$, and $\varepsilon$ mean the pressure, the rest mass
density, and the total energy density, respectively. Here,  $N$ and
$K$ are the polytropic index and the polytropic constant, respectively.
In the present study, numerical solutions of the equilibrium states are
constructed with the numerical method devised by Komatsu, Eriguchi \&
Hachisu (1989).

\subsection{Pulsation Equations in the Cowling Approximation}

In general relativity, not only the fluid but also the spacetime is
dynamical. We therefore have to solve linearized
Einstein equation in order to determine exact oscillation frequencies of
relativistic stars. When we consider oscillations of rapidly rotating
stars within the framework of general relativity, it is too difficult to
solve the linearized Einstein equation. No one has succeeded in obtaining a
general method for treating pulsations in rapidly rotating relativistic
stars. In order to avoid this difficulty, we employ a relativistic version
of the Cowling approximation for determining the $r$-mode oscillation
frequencies of rapidly rotating stars. We therefore neglect all the metric
perturbations and solve only the linearized continuity equation and the
linearized Euler equations, assuming the perturbations to be adiabatic.
The dynamics of the fluid can then be described in terms of the rest mass
density perturbation, $\delta\rho$, and the perturbations of the fluid
velocity, $\delta u^\mu$, where $\delta q$ denotes the Eulerian perturbation
of a physical quantity $q$.  Since perturbations of axisymmetric stars in
stationary states are considered, we assume that the perturbations have
harmonic time-dependence and $\varphi$-dependence given by
$e^{- i\sigma t+im\varphi}$, where $m$ is an integer and $\sigma$ is an
oscillation frequency measured by an inertial observer at spatial infinity.
The pulsation equations are given as follows.

For adiabatic perturbations,
\begin{eqnarray}
{\delta p\over p}=\left(1+{1\over N}\right)\,{\delta\rho\over\rho}\,,
\end{eqnarray}
where $\delta p$ is the Euler perturbation of the pressure and
we have assumed a barotropic relation with the same index
for both equilibrium states
and for pulsations,
i.e. the adiabatic index, $\Gamma$, is taken to be
$\Gamma=1+N^{-1}$. Note that due to the barotropic equation of state,
the star is neutrally stable against convection.

The perturbed continuity equation is expressed as
\begin{eqnarray}
&   &
   {u^t \over \rho} (\sigma - m \Omega) \delta \rho
   - {\partial \delta{\tilde u}^r \over \partial r}
   - {\partial \delta{\tilde u}^{\theta} \over \partial \theta} \nonumber\\
& - &
 \left\{ {2 \over r} + {1 \over \rho} {\partial \rho \over \partial r}
    + {\partial \over \partial r} (2 \alpha + \beta + \nu) \right\}
       \delta {\tilde u}^r \nonumber \\
& - &
 \left\{ \cot \theta + {1 \over \rho} {\partial \rho \over \partial \theta}
    + {\partial \over \partial \theta} (2 \alpha + \beta + \nu) \right\}
      \delta {\tilde u}^{\theta} \nonumber \\
& + &
 \left\{ \sigma F - m \right\} \delta u^{\varphi} = 0 \ ,
\end{eqnarray}
\begin{eqnarray}
 F \equiv {v^2  \over \Omega - \omega + v^2 \omega}
 \ ,
\end{eqnarray}
where $u^t$, $\delta u^r=i\delta{\tilde u}^r$,
$\delta u^{\theta}=i\delta{\tilde u}^\theta$,
and $\delta u^{\varphi}$, are the $t$-component of the four velocity,
the Euler perturbations of the r-component, $\theta$-component, and
$\varphi$-component of the four velocity, respectively.
The perturbed equations for the $r$-component, $\theta$-component
and  $\varphi$-component of the equations of motion
are:
\begin{eqnarray}
&   &
 {1 \over \varepsilon + p} {e^{-2 \alpha} \over u^t}
  {\partial \delta p \over \partial r}
  - {1 \over (\varepsilon + p)^2} {e^{-2 \alpha} \over u^t}
         {\partial p \over \partial r} (\delta \varepsilon + \delta p)
   \nonumber \\
& + &
 (\sigma - m \Omega) \delta {\tilde u}^r \nonumber \\
& - &
  e^{2 \beta - 2 \alpha} r^2 \sin^2 \theta (\Omega - \omega)
    {\partial \ln F \over \partial r} \delta u^{\varphi} = 0 \ ,
\end{eqnarray}
\begin{eqnarray}
&   &
  {1 \over \varepsilon + p} {e^{-2 \alpha} \over r^2 u^t}
    {\partial \delta p \over \partial \theta}
  - {1 \over (\varepsilon + p)^2} {e^{-2 \alpha} \over r^2 u^t}
         {\partial p \over \partial \theta} (\delta \varepsilon + \delta p)
     \nonumber \\
& + &
 (\sigma - m \Omega) \delta {\tilde u}^{\theta} \nonumber \\
& - &
   e^{2 \beta - 2 \alpha} \sin^2 \theta (\Omega - \omega)
     {\partial \ln F \over \partial \theta} \delta u^{\varphi} = 0 \ ,
\end{eqnarray}
\begin{eqnarray}
&    &
  {u^t \over \varepsilon + p} \left\{ (\sigma - m \omega)
    - {\Omega - \omega  \over v^2} m \right\} \delta p
     \nonumber \\
& - &
   {\partial \ln H \over \partial r} \delta {\tilde u}^r
  - {\partial \ln H \over \partial \theta}  \delta {\tilde u}^{\theta}
   \nonumber \\
& + &
   {\sigma - m \Omega \over \Omega - \omega + \omega v^2}
     \delta u^{\varphi} = 0 \ ,
\end{eqnarray}
\begin{eqnarray}
  H \equiv  {1 \over \Omega - \omega} { v^2 \over  1- v^2} \ ,
\end{eqnarray}
where $\delta \varepsilon$ is the Euler perturbation of the energy density.

Physically acceptable solutions of pulsation equations have to satisfy
boundary conditions on the rotation axis and on the stellar surface. The
regularity conditions are imposed on the rotation axis, i.e. all the
eigenfunctions have to vanish on the rotation axis if non-axisymmetric
perturbations are considered. The surface boundary condition is given by
\begin{eqnarray}
-i \gamma (\sigma-m\Omega)\Delta p=-i \gamma \,(\sigma-m\Omega)\delta p+
\delta u^\mu\,{\partial p\over\partial x^\mu} = 0 \,,
\end{eqnarray}
where $\Delta q$ means the Lagrangian perturbation of a physical quantity $q$,
which is related to the Eulerian perturbation $\delta q$ through the relation
$\Delta q=\delta q+{\cal L}_\xi q$. Here ${\cal L}_\xi$ denotes the
Lie derivative along the Lagrangian displacement vector $\xi^\mu$.

As in the Newtonian models (see e.g. Karino et al. 2000), we introduce a
surface-fitted coordinate system as follows:
\begin{eqnarray}
  r_* \equiv {r \over r_s(\theta)} \ ,  \qquad \theta_* \equiv \theta \ ,
\end{eqnarray}
where $r = r_s(\theta)$ denotes the surface shape of the unperturbed
configuration. In the actual numerical computations, we make use of
equidistant mesh points both in the radial and angular
directions ($0\le r_* \le 1$; $0\le\theta_*\le\pi/2$). Note that it is
sufficient to consider the angular range $0\le\theta_*\le\pi/2$ to
determine oscillation modes of rotating axisymmetric equilibrium stars,
because eigenfunctions of the stars must have reflection symmetry or
reflection anti-symmetry with respect to the equatorial plane. For the
$r$-mode oscillations we consider, the eigenfunctions are reflection
anti-symmetric.
In the present investigation, we take the number of mesh points to be mostly
$(r_*\times\theta_*)=(101 \times 70)$ and sometimes
$(r_*\times\theta_*)=(121 \times 101)$. In order to obtain numerical solutions
of the $r$-mode oscillations, pulsation equations described before are
transformed to a new surface-fitted coordinate system and
discretized on the mesh points, and are cast into coupled non-linear
algebraic equations in terms of the oscillation frequency $\sigma$ and
the discretized eigenfunctions, which are solved with the Newton-Raphson
iteration scheme. The numerical procedure we use is based on that of Yoshida
and Eriguchi (1997) and is a generalization of that of Karino et al. (2000).

\section{Numerical Results}

\begin{table}
\centering\caption{Equilibrium sequences for which the $r$-mode oscillations
have been analyzed. Here, the relativistic
factors $M/R$ shown in this table are evaluated in the non-rotation limit.
Here $M$ and $R$ are the gravitational mass and the circumferential radius
of the equilibrium models.
Therefore, the values of $M/R$ along the equilibrium sequences
are not constant but their variations are not large (typically a few percent
for $N=0.5$ and 10\% for $N=1$). $p_c$ and $\varepsilon_c$ are the pressure
and the energy density at the center of the star.}
\begin{tabular}{*{4}{c}}
\hline
 Sequence  & $N$ & $M/R$ & $p_c/\varepsilon_c$\\[0.5ex]
\hline
a & 0.5 & 0.1 & 0.06180\\[0.5ex]
b &     & 0.2 & 0.1744\\[0.5ex]
c & 1.0 & 0.1 & 0.06614\\[0.5ex]
d &     & 0.2 & 0.2016\\[0.5ex]
\hline \label{eq model}
\end{tabular}
\end{table}
\begin{table}
\centering\caption{$R$-mode frequency in the limit of no rotation.
The results of Yoshida and Lee (2002) in which the relativistic
Cowling approximation is used are also shown. The results of the
full problem are obtained by solving the full equations for the r-mode
oscillations in the slow rotation approximation. The quantities $\omega_c$
and $\omega_{\rm eq}$ are the values of the metric function $\omega$
at the center and the surface, respectively.}
\begin{tabular}{*{7}{c}}
\hline
 Method & $N$ & $M/R$ & $p_c/\varepsilon_c$ & $\omega_{\rm eq}/\Omega$ &
$\omega_c/\Omega$ & $\sigma/\Omega$ \\[0.5ex]
\hline
Present        & 0.5 & 0.10 & 0.06180 & 0.073 & 0.236 & 1.410\\[0.5ex]
Yoshida \& Lee & 0.5 & 0.10 &  ---    &  ---  &  ---  & 1.410\\[0.5ex]
Full           & 0.5 & 0.10 &  ---    &  ---  &  ---  & 1.379\\[0.5ex]
Present        & 0.5 & 0.20 & 0.1744  & 0.167 & 0.478 & 1.502\\[0.5ex]
Yoshida \& Lee & 0.5 & 0.20 &  ---    &  ---  &  ---  & 1.503\\[0.5ex]
Full           & 0.5 & 0.20 &  ---    &  ---  &  ---  & 1.443\\[0.5ex]
Present        & 1.0 & 0.10 & 0.06614 & 0.059 & 0.274 & 1.410\\[0.5ex]
Yoshida \& Lee & 1.0 & 0.10 &  ---    &  ---  &  ---  & 1.410\\[0.5ex]
Full           & 1.0 & 0.10 &  ---    &  ---  &  ---  & 1.380\\[0.5ex]
Present        & 1.0 & 0.20 & 0.2016  & 0.139 & 0.556 & 1.505\\[0.5ex]
Yoshida \& Lee & 1.0 & 0.20 &  ---    &  ---  &  ---  & 1.506\\[0.5ex]
Full           & 1.0 & 0.20 &  ---    &  ---  &  ---  & 1.453\\[0.5ex]
\hline \label{eq model2}
\end{tabular}
\end{table}

Since the $m=2$ $r$-mode is the most unstable against gravitational
radiation reactions among the inertial modes (Lockitch \& Friedman 1999;
Yoshida \& Lee 2000a; Yoshida \& Lee 2000b), it would be the most
important mode in the spin evolution of neutron stars. Therefore,  in the
present study, we concentrate on the $r$-mode oscillations associated with
$m=2$.  There are neither overtone modes nor modes associated with $l\ne |m|$
in the non-rotation limit, where $l$ is an angular eigenvalue of the
spherical harmonic function, for the $r$-mode oscillations of  barotropic
stars (Provost, Berthomieu \& Rocca 1981; Lockitch \& Friedman 1999).
Thus, only fundamental $r$-modes with $l = |m|$, whose eigenfunctions have
no node in the radial direction, are calculated. By decreasing the axis ratio
$r_p/r_e$ of the equilibrium star from $r_p/r_e=1$, where $r_p$ and $r_e$
are the polar radius and the equatorial radius of the star, respectively,
we obtain equilibrium configurations from a nonrotating spherical star to
a deformed star at the mass-shedding limit by the Komatsu, Eriguchi, and
Hachisu numerical scheme (Komatsu, Eriguchi \& Hachisu 1989, KEH). Along an
equilibrium sequence, the ratio of the pressure to the total energy density
at the center of the star, $p_c/\varepsilon_c$, the polytropic index, $N$, are
kept constant. As another rotation parameter, in addition to $r_p/r_e$, we
use the ratio of the rotational energy ($T$) to the absolute value of the
gravitational energy ($W$), $T/|W|$ (for the definition of $T/|W|$, see,
 Komatsu, Eriguchi \& Hachisu 1989). Physical quantities characterizing
the stars along an equilibrium sequence are tabulated in Table 1.

In order to check our numerical code, we have computed several $r$-mode
oscillations for slowly rotating stars and extrapolated the $r$-mode
frequencies to the limit of $\Omega\to 0$ from the obtained frequencies. The
results in the non-rotating limit are compared with the solutions obtained
by Yoshida and Lee's numerical scheme, in which all oscillation modes of
slowly rotating relativistic stars have been computed under the relativistic
Cowling approximation (Yoshida \& Lee 2002). The results summarized in
Table. 2 show that oscillation frequencies obtained by two different
methods are in excellent agreement and that the numerical scheme developed
in the present study works quite well. In this table, $\omega_c$ and
$\omega_{\rm eq}$ denote the values of the metric function
$\omega$ at the center and the surface of the spherical star, respectively.

In Table. 2, the results for the r-mode oscillations of the full equations
including the metric perturbations are also shown. These results are obtained
by using the formalism proposed  by Lockitch et al. (2001)
(see, also, Yoshida \& Lee 2003).  Although it
was very difficult to obtain the smooth eigenfunction for the model with
$N = 1.0$ and $M/R = 0.2$ in the full analysis, the obtained eigenfrequency
for that model should not be far from the true value. As seen from this
table, the relativistic Cowling approximation can give very good
approximate values for the oscillation frequencies to within a few percent.

Stergioulas \& Font (2001) have extracted an $r$-mode-like
characteristic frequency from their general relativistic simulations.
They performed hydrodynamical simulations using the relativistic
Cowling approximation where the spacetime geometry was fixed during
the time evolution of the fluid. Therefore, the present results can be
directly compared with their results. We have obtained a frequency of
$1.05$ kHz for the $m=2$ $r$-mode oscillation of an $N=1$ rotating
polytrope similar to that of Stergioulas \& Font, whose gravitational mass,
equatorial circumferential radius, and spin period are $M=1.63 M_\odot$,
$R=17.17$ km, and $P=1.25$ ms, respectively. Our result is in good agreement 
with the oscillation frequency of the $r$-mode obtained by 
Stergioulas \& Font, i.e. $1.03$ kHz.

In Figures 1 and 2, the ratio of the $r$-mode frequency to the angular
velocity, $\sigma/\Omega$, of rotating stars is shown as a function of
$T/|W|$ for four equilibrium sequences listed in Table 1. Note that in the
present study the frequency measured by an inertial observer at spatial
infinity is shown; the corotating frequency $\sigma_c$ used
in many papers is related to $\sigma$ by $\sigma_c = \sigma - m \Omega$.
Figures 1 and 2
show results for polytropes with $N=0.5$ (sequence $a$ and sequence $b$)
and $N=1.0$ (sequence $c$ and sequence $d$), respectively.  From these
figures, it is found that the scaled eigenfrequencies of the $r$-mode
oscillations, $\sigma/\Omega$, are decreasing functions of $T/|W|$ and are
given as nearly linear functions of $T/|W|$ except for configurations in the
mass-shedding states. Similar features have
been found in the Newtonian $r$-mode oscillations of rapidly rotating stars
(Karino, Yoshida \& Eriguchi 2001). It is important to note that this simple
relation between $\sigma/\Omega$ and $T/|W|$ is preserved even for highly
relativistic stars. Since the oscillation frequencies $\sigma/\Omega$ are
nicely approximated by linear functions of $T/|W|$ except for the region near
the mass-shedding limit, it is meaningful to derive approximation formulas for
$\sigma/\Omega$ versus $T/|W|$, which are given by
\begin{equation}
{\sigma\over\Omega}\approx\left\{
\begin{array}{ll}
1.41-1.85\,{T\over |W|} & \mbox{for sequence $a$}\\
1.50-1.23\,{T\over |W|} & \mbox{for sequence $b$}\\
1.41-1.95\,{T\over |W|} & \mbox{for sequence $c$}\\
1.51-1.36\,{T\over |W|} & \mbox{for sequence $d$}\,.
\end{array}
\right.
\label{fre-approx}
\end{equation}
Here, after discarding data near the mass-shedding limit, we apply a
least-square-fit to the scaled frequencies as functions
of $T/|W|$. Comparing the coefficients appearing in these formula
(\ref{fre-approx}) for different sequences, we can see that the $r$-mode
frequencies of relativistic stars do not strongly depend on the stiffness of
the equation of state, but they are sensitive to the relativistic
factor $M/R$ and to the rotation rate. Similar strong dependence of the
relativistic factor $M/R$ on the oscillation frequencies of the $r$-mode
oscillations has been already observed in the $r$-mode oscillations of slowly
rotating relativistic stars (e.g., Kojima 1998; Lockitch, Andersson
\& Friedman 2001; Yoshida 2001; Ruoff \& Kokkotas 2001; Yoshida \& Lee 2002;
Lockitch, Friedman \& Andersson 2003).  As discussed by Kojima (1998),
frequencies of the $r$-mode oscillations in the limit of no
rotation can be approximated by a certain average of the local
oscillation frequency of the $r$-mode oscillation, namely
\begin{equation}
{\sigma\over\Omega}={1\over |m|+1}\,\left\{(|m|+2)(|m|-1)+
2\,{\omega\over\Omega} \right\}\,.
\label{localw}
\end{equation}
In equation (\ref{localw}), the relativistic effect appears only in the
second term in the braces, which is always positive for non-pathologically
rotating stars.  Therefore, the frequency of the $r$-mode oscillation tends
to increase as the relativistic factor $M/R$ is increased because the
dragging of inertial frame becomes significant for highly relativistic
stars as seen from Table 2. This property is consistent with our numerical
results as shown in Figures 1 and 2.

In Figures 3 and 4, typical distributions of velocity perturbations
$\delta v_r$ and $\delta v_{\theta}$  are displayed for the $r$-modes of
slowly and rapidly rotating stars, respectively. Here we define,
\footnote{Note that these are not exactly the components of 3 velocity on the
orthonormal frame which is naturally defined on our spacetime.}

\begin{eqnarray}
\delta v_r \equiv \delta {\tilde u}^r \ ,
\end{eqnarray}
\begin{eqnarray}
\delta v_{\theta} \equiv r \delta {\tilde u}^{\theta} \ .
\end{eqnarray}

The equilibrium models shown in the figures are taken from
the sequence $b$ in Table 1. The rotation parameters of the stars shown
in Figures 3 and 4 are chosen as $r_p/r_e=0.96$ $(T/|W|=0.01248)$ and
$r_p/r_e=0.70$ $(T/|W|=0.1035)$, respectively. In each figure, the
amplitudes of the eigenfunctions are shown against the surface-fitted radial
coordinate $r_*$ ($0\le r_*\le 1$) for five different constant values
of $\theta_*$, whose values are given in the figures. Figures 3 and 4 show
that the relativistic $r$-mode oscillations of rotating
stars are very similar to the corresponding Newtonian modes.
Moreover, comparing Figure 3 with Figure 4, we observe that the basic
properties of the $r$-mode oscillations of relativistic stars depend only
weakly on the magnitude of the rotation.
However, it is important
to emphasize the following properties; the horizontal motion of the fluid
due to the $r$-mode oscillations is more concentrated near the surface of
the star for a rapidly rotating star than for a slowly rotating star.
Similar behaviors are observed in the $r$-mode oscillations of Newtonian
stars (Karino et al. 2000).

For slowly rotating stars, the $r$-mode oscillations of Newtonian stars
do not excite the fluid motion in the radial direction, while in the
relativistic star, the fluid element can move in this direction due to
the effect of general relativity (Lockitch, Andersson \& Friedman 2000).
This property can be confirmed in Figure 3, which shows that the
perturbations of the radial velocity $\delta v_r$ for a slowly rotating
relativistic star have non-vanishing amplitudes. On the other hand,
as already shown
in Newtonian studies (Provost, Berthomieu \& Rocca 1981; Saio 1982),
the amplitude of $\delta v_r$ for the $r$-mode oscillations becomes
large as a rotation parameter $T/|W|$ of the equilibrium star is
increased. This fact is also observed in Figure 4.
In summary, our results for modal properties are consistent with previous
results that properties of pure-axial perturbations for the $r$-mode
oscillations cease to appear for a highly relativistic star and/or
for a rapidly rotating star.

\section{Discussion and Summary}

\subsection{Discussion}

It is the most crucial thing to know how far we could rely on the results
obtained by using the Cowling approximation. Since no one has succeeded in
developing a formulation for the full analysis of $r$-mode oscillations of
rapidly rotating stars in general relativity, what we can do is to estimate
the accuracy by considering the results obtained for the Newtonian
configurations and those for slowly rotating general relativistic stars.

The frequencies of $m = 2$ $r$-mode oscillations for the Newtonian polytropes
are shown in Table 3. In this table, the ratios of the $r$-mode oscillation
frequency to the angular velocity in the full problem and in the Cowling
approximation are tabulated both for slowly rotating polytropes and
for rapidly rotating polytropes with the polytropic indices $N = 0.5$ and $1$.
As seen from this table, the values of the Cowling approximation are
to within a several percent from the exact values. Since there occurs no
density perturbation in the $r$-mode oscillation of spherical Newtonian
stars, the Cowling approximation gives very accurate values for slowly
rotating Newtonian configurations irrespective of the polytropic indices.
This kind of behavior obtained by using the Newtonian Cowling approximation
is not directly applied to the general relativistic configurations
because the basic equations in general relativistic oscillations are
not the same as those in the Newtonian models. However,
it should be noted that the results of the scaled eigenfrequencies
$\sigma/\Omega$ obtained by the Cowling approximation are located
within a several percent from those of the full problem.
Together with the errors for general relativistic spherical stars
discussed in the previous section, we may as well believe that the results
obtained in this paper for the $r$-mode oscillations of
rapidly rotating general relativistic polytropes would be also
well approximated ones.

\begin{table}
\centering\caption{$R$-mode frequencies of uniformly rotating
Newtonian polytropes. The results for the full problem and those
obtained by the Cowling approximation are tabulated for $N = 0.5$ and $1$
polytropes. Here $r_p/r_e$ is the axis ratio of the rotating configurations,
where $r_p$ and $r_e$ are the polar radius and the equatorial radius,
respectively }
\begin{tabular}{*{4}{c}}
\hline
 Method & $N$ & $r_p/r_e$ & $\sigma/\Omega$ \\[0.5ex]
\hline
Full     & 0.5 &  0.98 & 1.320\\[0.5ex]
Cowling  & 0.5 &  0.98 & 1.319\\[0.5ex]
Full     & 0.5 &  0.67 & 1.099\\[0.5ex]
Cowling  & 0.5 &  0.67 & 1.050\\[0.5ex]
Full     & 1.0 &  0.98 & 1.323\\[0.5ex]
Cowling  & 1.0 &  0.98 & 1.321\\[0.5ex]
Full     & 1.0 &  0.67 & 1.159\\[0.5ex]
Cowling  & 1.0 &  0.67 & 1.116\\[0.5ex]
\hline \label{eq model3}
\end{tabular}
\end{table}

In general, the master equations for oscillations of a stationary
and axisymmetric rotating
star can be reduced to a single two-dimensional
second-order partial differential equation as shown below. However, the type
of the reduced second order partial differential equation cannot be known
beforehand because it depends on the oscillation frequency of the mode (see,
e.g., Balbinski 1985;
Skinner \& Lindblom 1996; Lindblom 1997).  In other words, it implies
that the type of the equation cannot be determined before some solutions of
the modes are obtained. This is an important fact because the
properness of the {\it boundary value problem} for the partial differential
equations is deeply related to the type of the differential equation.
If the equation is of the elliptic type, the boundary value problem is a
properly posed one. However, if it is hyperbolic, one might need to
devise a proper treatment of the boundary value problem for the
hyperbolic equations.

In our relativistic Cowling problem, we can derive the following second order
partial differential equation in terms of the pressure perturbation and
consider its type. From the basic equations (7) and (11)-(13), we can obtain
\begin{eqnarray}
A {\partial^2 \delta p \over \partial r^2}
+ B {\partial^2 \delta p \over \partial r \partial \theta}
+ C {\partial^2 \delta p \over \partial \theta^2} +
f({\partial \delta p \over \partial r}, {\partial \delta p \over
\partial \theta}, \delta p, r, \theta) = 0 \ ,
\label{formofmastereq}
\end{eqnarray}
\begin{eqnarray}
 A &\equiv&  {1 \over \varepsilon + p} {e^{-2 \alpha} \over u^t}
  {1 \over D (\sigma - m \Omega)^2}
\left\{ {(\sigma - m \Omega)^2 \over \Omega - \omega + \omega v^2}
\right. \nonumber\\
   &  -   &
\left. {v^2 \over \Omega - \omega} {1 \over r^2}
{\partial \ln F \over \partial \theta} {\partial \ln H \over \partial \theta}
\right\} \ ,
\end{eqnarray}
\begin{eqnarray}
 B & \equiv & {1 \over \varepsilon + p} {e^{-2 \alpha} \over u^t}
  {1 \over D (\sigma - m \Omega)^2} {1 \over r^2}
           {v^2 \over \Omega - \omega} \nonumber\\
   & \times  &
\left\{
{\partial \ln F \over \partial r} {\partial \ln H \over \partial \theta} +
{\partial \ln F \over \partial \theta} {\partial \ln H \over \partial r}
\right\} \ ,
\end{eqnarray}
\begin{eqnarray}
 C & \equiv & {1 \over \varepsilon + p} {e^{-2 \alpha} \over u^t}
  {1 \over D (\sigma - m \Omega)^2} {1 \over r^2} \nonumber\\
   & \times &
\left\{ {(\sigma - m \Omega)^2 \over \Omega - \omega + \omega v^2}
  -{v^2 \over \Omega - \omega}
{\partial \ln F \over \partial r} {\partial \ln H \over \partial r}
\right\} \ ,
\end{eqnarray}
\begin{eqnarray}
 D & \equiv &  {1 \over \sigma - m \Omega}
\left\{ {(\sigma - m \Omega)^2 \over \Omega - \omega + \omega v^2}
\right. \nonumber\\
 &  -  &
\left. {v^2 e^{2\nu-2\alpha}\over \Omega - \omega}
\left({\partial \ln F \over \partial r} {\partial \ln H \over \partial r}
+ \frac{1}{r^2} {\partial \ln F \over \partial \theta}
{\partial \ln H \over \partial \theta} \right)
\right\} \ .
\end{eqnarray}
where $f$ is a function of the first order derivative of $\delta p$,
$\delta p$ itself, and coordinates $(r, \theta)$ whose form is not essential
for the following discussion (see, also, Ipser \& Lindblom 1992).

The type of the above partial differential equation (21) at each mesh point
can be determined by the following quantity $Q$:
\begin{eqnarray}
  Q \equiv B^2 - 4 A C \ .
\end{eqnarray}
For a uniformly rotating Newtonian barotropic star, the quantity $Q$ is
given by
\begin{eqnarray}
  Q = {16 \over r^2} {1 \over 4 \Omega^2-(\sigma-m\Omega)^2} \ .
\end{eqnarray}
Accordingly, the master equation in the Newtonian limit is
hyperbolic (elliptic) if the frequency satisfies
$4\Omega^2>(\sigma-m\Omega)^2$ ($4\Omega^2<(\sigma-m\Omega)^2$).
It is important to note that oscillation frequencies of all inertial modes
of uniformly rotating Newtonian stars that have been obtained so far are in
the range of $-2\Omega<\sigma-m\Omega<2\Omega$, implying that
Eq.(\ref{formofmastereq}) is the hyperbolic (cf. Greenspan 1969).

As for all general relativistic r-mode oscillations obtained in this paper,
values of the quantity $Q$ are also positive at every point inside the star.
It follows that the equation is again hyperbolic.
Although there is no mathematical proof, therefore, it seems that solutions
of the inertial modes are basically given as solutions of partial differential
equations of the hyperbolic type. Note that the situation would be more
complicated if non-barotropic stars were considered. In non-barotropic stars,
the inertial modes behave like $g$-modes which are solutions of
an elliptic equation when the rotational velocities of the stars are
sufficiently slow (see, e.g., Yoshida \& Lee 200b).

\subsection{Summary}

In this paper we have developed a numerical scheme for obtaining the
$r$-mode oscillations of rapidly rotating relativistic stars. In the present
scheme, we neglect all the metric perturbations and only take account of the
dynamics of the fluid motion in the fixed background spacetime of the star
(the relativistic Cowling approximation). We also assume stars to be
barotropic under the assumption of adiabatic oscillations. Our numerical
scheme is based on the Yoshida-Eriguchi formulation (Yoshida \& Eriguchi 1997)
and is a general relativistic extension of the
Karino-Yoshida-Yoshida-Eriguchi numerical scheme for determining
oscillations of rapidly rotating Newtonian stars (Karino et al. 2000). With
this new numerical scheme, the frequencies of the $r$-mode oscillations are
obtained as functions of the ratio of the rotational energy to the absolute
value of the gravitational energy $T/|W|$ along sequences of polytropic
equilibrium stars whose ratios of the pressure to the total energy density
at the center of the stars and the polytropic index are kept constant. It is
found that scaled oscillation frequencies $\sigma/\Omega$ are decreasing
functions of $T/|W|$ and are given as nearly linear functions of $T/|W|$. As
already found in studies on the relativistic $r$-mode oscillations of slowly
rotating stars, we observe that the $r$-mode oscillation frequencies are
sensitive to change of the relativistic factor of the stars but not to the
change of stiffness of the equation of state. The present results are
consistent with the previous results of the $r$-mode oscillations both for
rapidly rotating Newtonian stars and for slowly rotating relativistic stars.

It should be noted that three simplifications about the basic properties of
the relativistic $r$-mode oscillations of rapidly rotating general
relativistic stars have been employed in the present investigation;
namely, the use of (1) the relativistic Cowling approximation,
(2) the rigid rotation law, and (3) the barotropic equation of state.
Although it is known that the Cowling approximation can produce approximate
eigenfrequencies with acceptable errors and basic modal properties similar to
the exact ones for such oscillation modes as the fluid motion is dominating,
the exact treatment of full equations for obtaining oscillation modes of
rapidly rotating relativistic stars is required for understanding accurate
solutions for the oscillation modes and for exploring purely general
relativistic properties such as damping times of oscillations due to
gravitational radiation reactions. A great challenging problem to find a
method for obtaining and solving exact pulsation equations for rapidly
rotating relativistic stars has been left unsolved yet.

An extension of the present method to differentially rotating stars and/or
to non-barotropic stars would be worthwhile in the following context:
nascent neutron stars are expected to rotate differentially and the effect of
differential rotation could strongly affect the frequencies and the modal
properties of the $r$-mode oscillations (for Newtonian models, see e.g.
Karino et al. 2001). In order to have a correct scenario of spin
evolutions of young neutron stars due to the $r$-mode instability, therefore,
we have to know how the $r$-mode oscillations would be changed under the
effect of differential rotation. As for the $r$-mode oscillations of
non-barotropic stars, as mentioned in Section I, we do not yet know
whether the $r$-mode oscillations similar to those of Newtonian stars, whose
eigenfrequencies are discrete and eigenfunctions are regular, can exist
in relativistic non-barotropic stars. Future investigations on
the relativistic $r$-mode oscillations of rapidly rotating non-barotropic
stars in the Cowling approximation might give us useful clues for
understanding of the full $r$-mode oscillations of non-barotropic
relativistic stars.

\vskip 0.5cm
\noindent{\sl  Acknowledgments:}
{ We thank John Friedman for his comments on the manuscript.
This research was supported in part by the Grant-in-Aid for the 21st
Century COE ``Holistic Research and Education Center for Physics of
Self-organization Systems'' from the ministry of  Education, Science,
Sports, Technology, and Culture of Japan, and by the Grant-in-Aid
for Scientific Research (C) (14540244) from Japan Society for the
Promotion of Science. S'i.Y is supported by NSF Grant PHY-0071044.}

\newpage

\begin{figure}
\centering
\includegraphics[width=8cm]{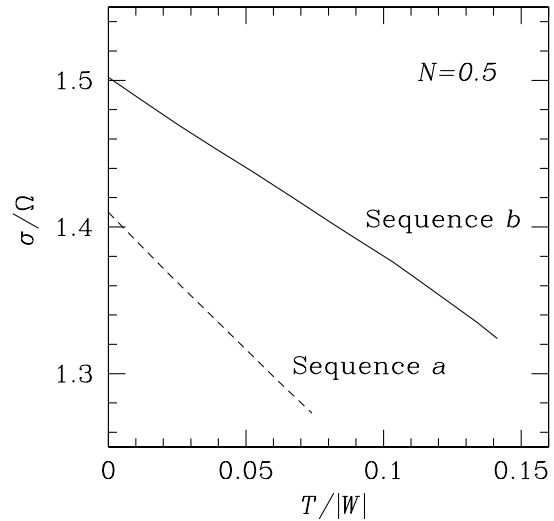}
\caption{Scaled frequencies $\sigma/\Omega$ of the $m = 2$ $r$-mode
oscillation of rotating stars with the polytropic index $N=0.5$
(sequence $a$ and sequence $b$) are plotted as functions of the ratio of the
rotational energy to the absolute value of the gravitational energy,
$T/|W|$. The frequencies of the $r$-mode oscillations for the sequence $a$
and the sequence $b$ are shown by dashed and solid curve, respectively.}
\end{figure}
\begin{figure}
\centering
\includegraphics[width=8cm]{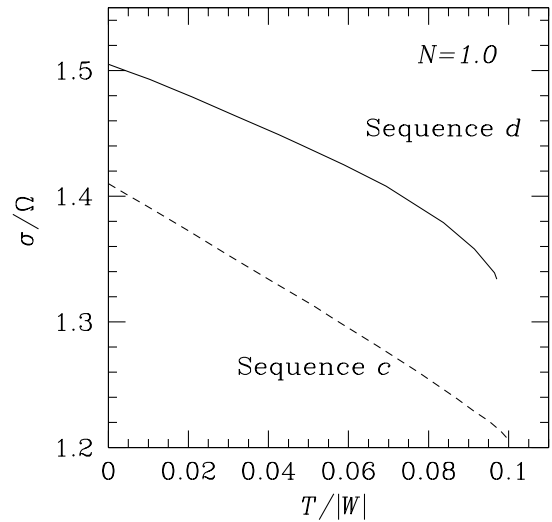}
\caption{Same as Figure 1 but for sequences $c$ (dashed curve) and
$d$ (solid curve).}
\end{figure}
\begin{figure}
\centering
\includegraphics[width=8cm]{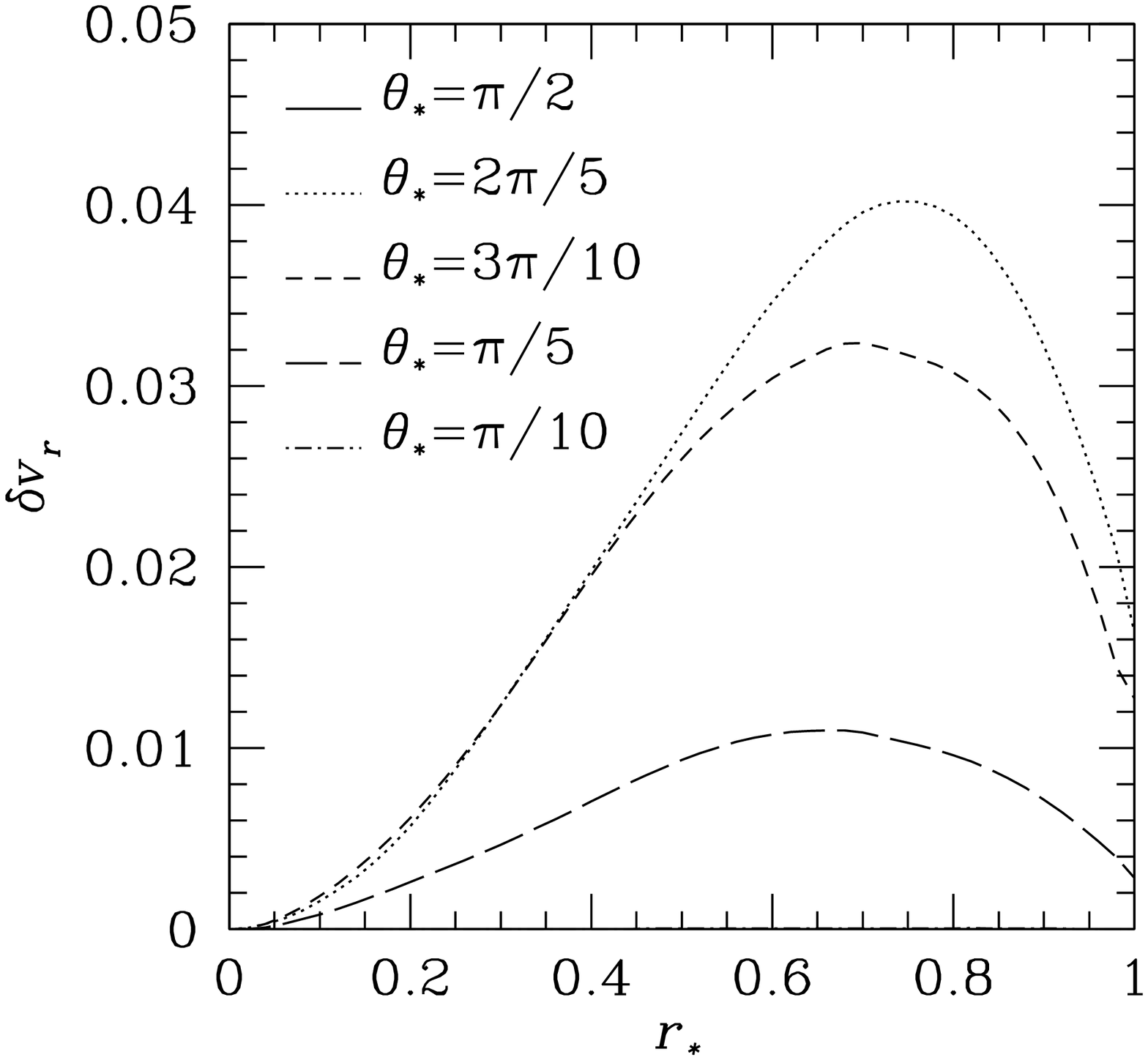}
\includegraphics[width=8cm]{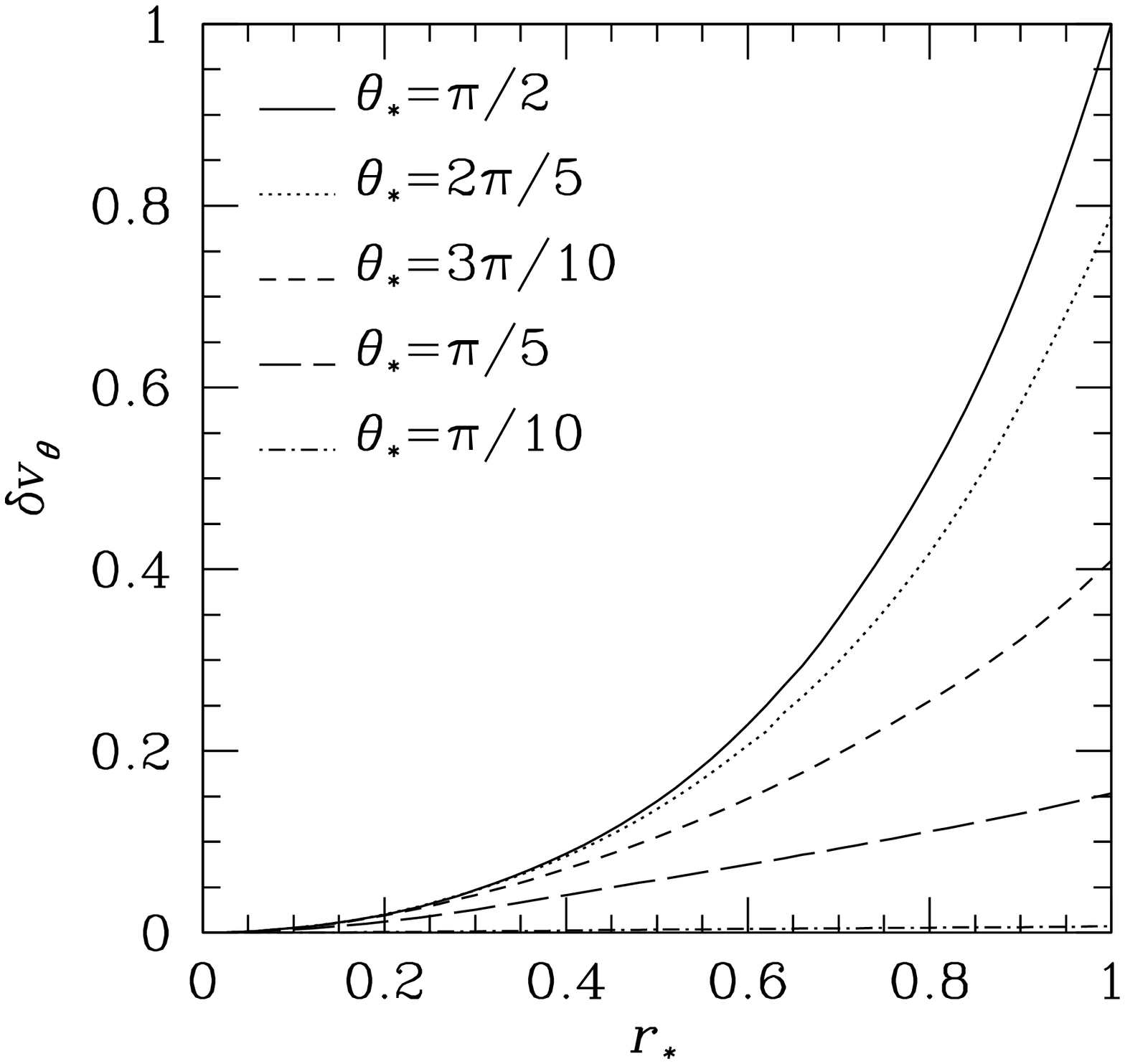}
\caption{Eigenfunctions $\delta v_r$ (top) and $\delta v_{\theta}$ (bottom)
of the $m = 2$ $r$-mode oscillation for a slowly rotating star.
The equilibrium configuration belongs to the sequence $b$ and
its rotation parameter is taken to be $r_p/r_e=0.96$ ($T/|W|=0.01248$).
Distributions of the eigenfunctions for several values of $\theta_*$
are displayed as functions of the surface fitted radial coordinate
$r_*$. Normalization of the eigenfunctions is taken so as to be
$\delta v_{\theta}=1$ at the stellar surface on the equator.
Note that $\delta v_r$ for $ \theta_*=\pi/10$ and $\pi/2$ are very small
and cannot be seen in this figure. }
\end{figure}
\begin{figure}
\centering
\includegraphics[width=8cm]{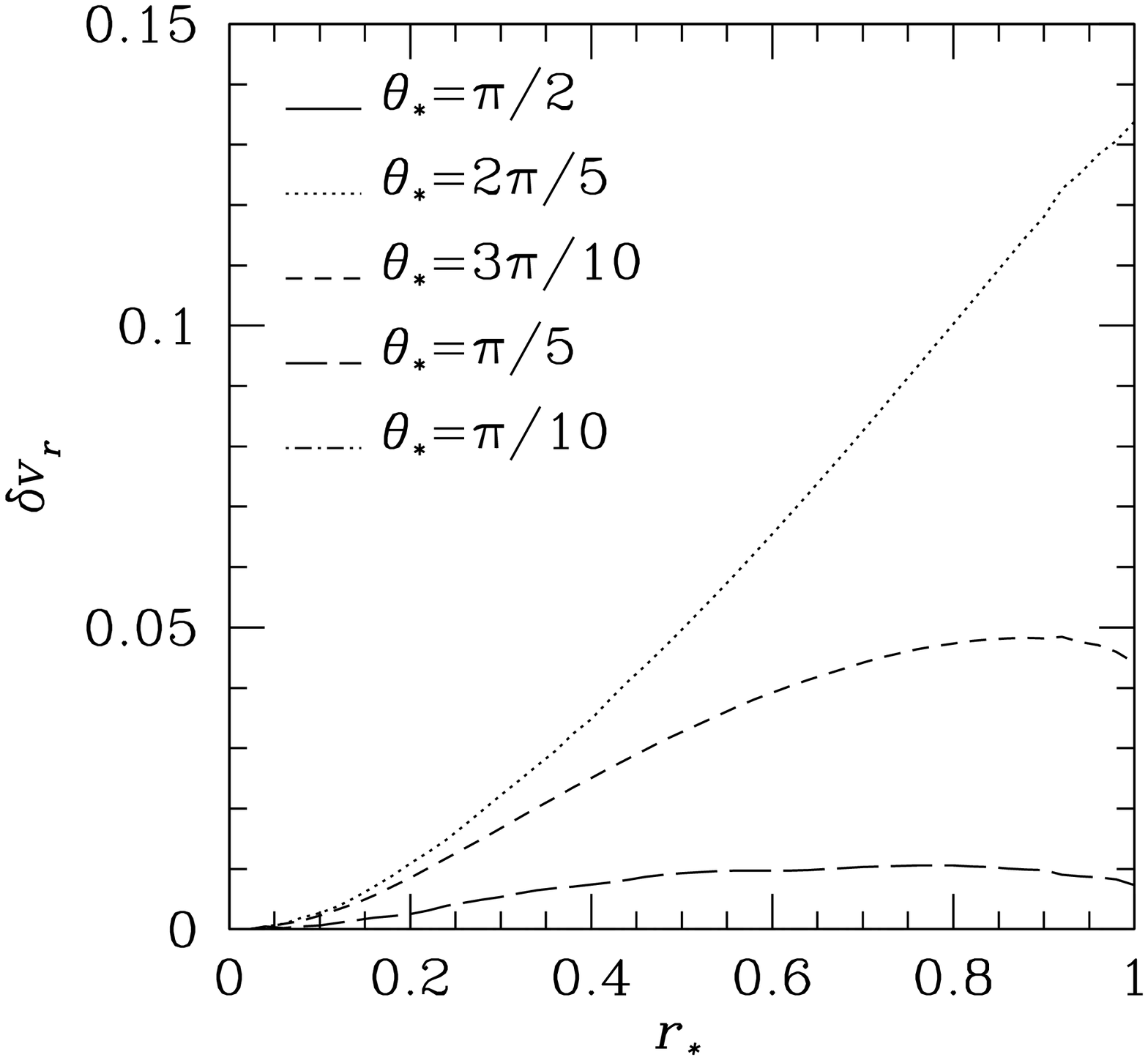}
\includegraphics[width=8cm]{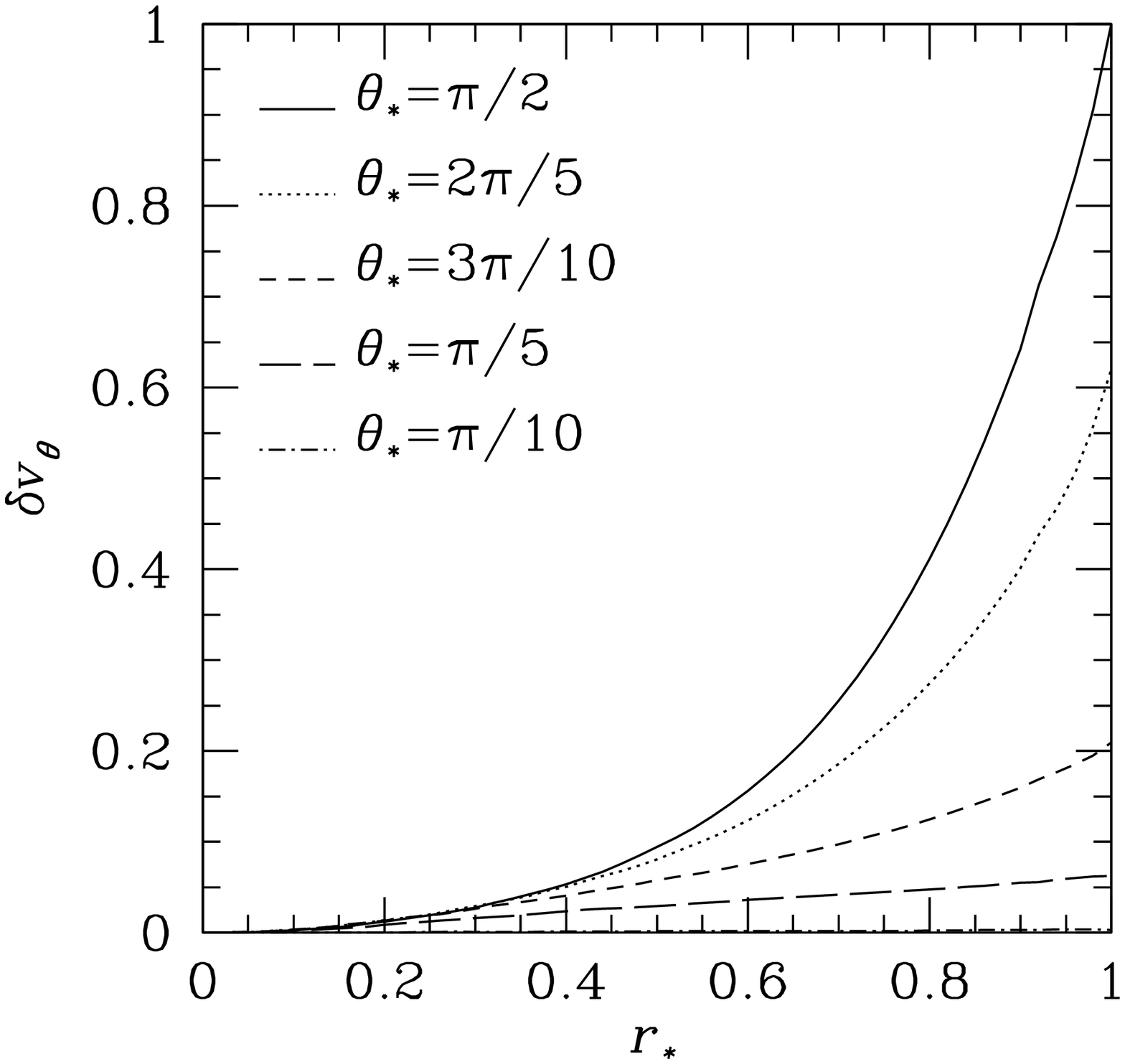}
\caption{Same as Figure 3 but for a rapidly rotating star
whose rotation parameter is taken to be $r_p/r_e=0.70$ ($T/|W|=0.1035$).
}
\end{figure}

\end{document}